**A critical reappraisal of predicting suicidal ideation using fMRI**

Timothy Verstynen[1] & Konrad Paul Körding[2]

1. Departments of Psychology, Carnegie Mellon Neuroscience Institute, and Biomedical Engineering, Carnegie Mellon University, Pittsburgh, PA

2. Departments of Bioengineering and Neuroscience, University of Pennsylvania, Philadelphia, PA, USA

**Correspondence**:
Timothy Verstynen
Email: timothyv@andrew.cmu.edu

## Abstract

For many psychiatric disorders, neuroimaging offers a potential for revolutionizing diagnosis, and potentially treatment, by providing access to preverbal mental processes. In their study "Machine learning of neural representations of suicide and emotion concepts identifies suicidal youth."[1], Just and colleagues report that a Naive Bayes classifier, trained on voxelwise fMRI responses in human participants during the presentation of words and concepts related to mortality, can predict whether an individual had reported having suicidal ideations with a classification accuracy of 91%. Here we report a reappraisal of the methods employed by the authors, including re-analysis of the same data set, that calls into question the accuracy of the authors findings. The analysis is a case study in the dangers of overfitting in machine learning.

**Main Text**

Unlike many areas of medicine, the fields of psychiatry and clinical psychology suffer from a critical lack of the ability to directly measure the internal processes that are the root of most psychiatric disorders[2]. Instead, these fields rely on indirect assessments, via verbal report or behavioral analyses, that can often be unreliable indicators of internal thoughts and experiences. Over the past few years, machine learning methods applied to functional neuroimaging data have presented a promising avenue for the field of computational psychiatry to potentially measure preverbal internal processes[3], offering hope for the development of neural biomarkers of psychiatric diseases.

In one such study, Just and colleagues[1], reported promising findings on a potential neural biomarker for suicidal ideation. The authors reported a 91% classification accuracy for predicting a participant's group membership (suicidal ideating individuals, N=17; non-ideating control; N=17), using leave one out cross validation (LOOCV) with a classifier trained on functional MRI (fMRI) responses to a list of words. Such a robust ability to identify individuals who are likely suicidally ideating based off of pre-verbal neural processes could revolutionize psychiatric approaches to suicide.

However, the procedures described in the original paper suggest several problems. First, the use of LOOCV can inflate the estimated classification accuracy, as well as overall Type-I error[4]. Second, and most importantly, the feature selection appears to have relied on the same data that is used in the final model evaluation. In the section titled "Identifying the most discriminable concepts and locations" of the Supplementary Information, the authors state that they used a forward stepwise selection procedure to identify the best combination of concepts (i.e., words) and locations (i.e., sphere of voxels from anatomically defined regions of interest) that maximized their model accuracy in predicting whether a participant was in the suicidal ideation or control group. According to the text, feature selection happened along two different dimensions: words and regions. For words, the authors only used data from 6 out of 30 words in their final model. The authors present no *a priori* reason for why this subset of words would be better at discriminating between groups. Therefore, we assume that this subset was determined solely by the described forward stepwise search process. We can not be entirely sure because the authors did not share all of their code. However, as we shall point out below, we have reasons to believe that we understand their approach reasonably well. For regions, the authors used multiple selection procedures for determining which clusters of voxels to include in their model, resulting in 5 out of approximately 25 (on average 25 based on group analysis) regions being included in the final model. First, the authors evaluate voxels based on a stability score of responses across trials. No information is provided for how this stability is quantified. For each fold of the cross validation procedure, the hold out test subject was not included in the voxel stability analysis.

Though it is not clear why, because voxel stability is already an independent measure from the classifier performance. Second, the authors selected the best subset of stable voxel clusters on each fold, separately for each group. On average there were 11 stable regions for the suicidal ideation group and 14 regions for the control group. The final analysis only included 2 from the suicidal ideation group and 3 from the control group, again presumably identified using the stepwise selection procedure. This is problematic, however, because group assignment is already influencing features included in the final classifier analysis. It is in the group subset that the forward stepwise search appears to have been applied.

Given the sample size and structure of the classification problem, we can see no way that a *consistent* set of features (i.e., one set of words and regions to serve as a biomarker across all subjects) can be identified without using data from all participants. This reflects what we are calling "feature hacking"[5], a form of circular inference[6] that contaminates the validity of out-of-sample validation tests. Feature hacking is the process of inflating model performance in cross validation tests by selecting the best subset of the features that maximizes performance of the hold out test set that is used as a benchmark for how well the classifier generalizes to predicting unseen data.

Using the code and data provided by the authors[7], we conducted a reanalysis of the feature selection process[8]. We started by simply attempting to replicate the deterministic feature selection method as described in the original manuscript, using a logistic regression classifier on group membership and a forward stepwise feature selection in three stages. First, we selected the subset of words that best distinguished the two groups using average response data from all stable regions (with stability determined excluding the data from the out of sample participant). Second, we selected the set of stable regions from the suicidal ideation group that best distinguished group membership using all 30 words. Finally, we selected the stable regions from the control group using all 30 words. Feature selection on regions was run separately for the two groups because this follows the logic of the original analysis. It is worth noting that this method still suffers from circular inference because all data are being used in the feature selection process. As highlighted above, the sample size is too small to enable a completely unbiased feature selection process.

Our analysis was unable to replicate the original feature set. We identified only one word, "vitality", one region from the patient group, left angular gyrus, and one region from the control group "left anterior cingulum". Only the latter region overlapped with the original set of features. Importantly, using these words and regions, the LOOCV classifier method from the original paper falls to 32% (see Table 1). Thus, using a standard forward stepwise selection procedure we were unable to either replicate the features or model accuracy reported in the original paper.

We next set out to see how much the feature selection process employed by the authors impacted the classifier performance. To start we re-ran the original classifier reported by Just et al. 2017, but removed feature selection along the two dimensions, words and regions, separately. These results are reported in Table 1. Removing feature selection on words, but including the same set of selected regions as used in the original paper, reduced classifier accuracy by 32%. Removing feature selection on the set of stable regions, while keeping the original 6 words used in the original paper, dropped classifier accuracy by 26%. Thus the classifier accuracy reported by Just et al., 2017 is highly sensitive to the unique set of words and regions used.

The only feature selection method employed in the original paper that does not suffer from circularity is the original selection of the stable voxel clusters (i.e., regions). Here stability was determined by excluding the held out test subject for each run of the LOOCV classifier. Thus, the only truly unbiased model that can be run on the data is one in which all words and all stable regions, for both groups, are used. This model returns a classification accuracy of 41%, well below chance and a full 50% below the accuracy reported by Just et al., 2017.

**Table 1.** Change in leave one out cross-validation (LOOCV) accuracy when different feature selection approaches are applied.

| Configuration | LOOCV Accuracy |
| --- | --- |
| Pre-selected features (words & regions) from Just et al. 2017. | 91% |
| Features from forward stepwise search using logistic regression. | 32% |
| All words, pre-selected regions | 59% |
| Pre-selected words, all regions | 65% |
| All words, all regions | 41% |

Using information from data in a validation set to determine the structure of a model leads to inflated estimates of performance. This can happen either by selecting the observations (e.g., only including the subset of participants that maximize validation set performance) or features (e.g., applying arbitrary transformations of variables based on validation set performance) based on information from what should be a protected part of the sample. Our reanalysis shows that the classification results reported by Just et al., 2017 are likely inflated due to the use of feature hacking. Our analysis clearly shows that the most conservative approach (using all words and all stable regions) yields classification performance that does not outperform chance. Our analysis suggests that the paper by Just and colleagues employs feature hacking. Without a more detailed description of the methods and independent evaluation of the feature selection process itself, we are forced to conclude that the reported ability to discriminate the suicidally ideating from non-ideating controls is not supported by the data reported in the original article.